\documentclass[aps,prl,twocolumn,superscriptaddress,nofootinbib]{revtex4-1}
\usepackage{amsmath}
\usepackage{amssymb}
\usepackage{graphicx}
\usepackage{mathrsfs}
\usepackage{ntheorem}
\usepackage{enumerate}
\usepackage{enumitem}
\usepackage{times,txfonts}
\usepackage{subfigure}
\usepackage{ulem}
\usepackage{titlesec}

\newcommand{\ket}[1]{|#1\rangle}
\newcommand{\bra}[1]{\langle #1|}
\newcommand{\Tr}{\mathrm{Tr}}

\newcommand{\abs}[1]{\lvert #1\rvert}

\def\CC{{\rm\kern.24em \vrule width.04em height1.46ex depth-.07ex \kern-.30em C}}
\def\RR{{\rm\kern.24em \vrule width.04em height1.46ex depth-.07ex
\kern-.30em R}}
\def\P{{\rm I\kern-.25em P}}

\begin{document}
\title{Estimating Coherence Measures from Limited Experimental Data Available}
\author{Da-Jian Zhang}
\affiliation{Department of Physics, Shandong University, Jinan 250100, China}
\affiliation{Department of Physics and Electronics, Shandong Normal University, Jinan 250014, China}
\author{C. L. Liu}
\affiliation{Department of Physics, Shandong University, Jinan 250100, China}
\author{Xiao-Dong Yu}
\affiliation{Department of Physics, Shandong University, Jinan 250100, China}
\author{D. M. Tong}
\email{tdm@sdu.edu.cn}
\affiliation{Department of Physics, Shandong University, Jinan 250100, China}

\date{\today}

\begin{abstract}
Quantifying coherence has received increasing attention, and considerable work has been directed towards finding coherence measures. While various coherence measures have been proposed in theory, an important issue following is how to estimate these coherence measures in experiments. This is a challenging task, since the state of a system is often unknown in practical applications and the accessible measurements in a real experiment are typically limited. In this Letter, we put forward an approach to estimate coherence measures of an unknown state from any limited experimental data available. Our approach is not only applicable to coherence measures but can be extended to other resource measures.
\end{abstract}

\maketitle
Coherence is a fundamental feature of quantum mechanics, describing the capability of a state to exhibit quantum interference phenomena. It plays a central role in physics, as it enables applications that are impossible within classical mechanics or ray optics \cite{Nielsen2010}. Its applications range from quantum computation \cite{Shor1997,Grover1997} to quantum cryptography \cite{Grosshans2001,Grosshans2003} and quantum metrology \cite{Giovannetti2004,Demkowicz2014,Giovannetti2011}.

A hot topic on coherence is its quantification, which has attracted a growing interest due to the development of quantum information science \cite{Aberg2006,Gour2008,Baumgratz2014,Marvian2014,Levi2014,Yu201602, Aberg2014,Streltsov2015,Bromley2015,Ma2015,Sun2015,Winter2016,Bagan2016,Napoli2016,
Radhakrishnan2016,JMa2016,Streltsov201602,Chitambar201601,Chitambar201603,Chitambar201602,
Yu201601,Yu201602,Long2016,Fan2016,Hu2016,Guo2017,Marvian2014,Du2015,Zanardi2017,Streltsov201603,Fan2017}.
The first rigorous framework for quantifying coherence is based on incoherent operations, put forward by Baumgratz, Cramer, and Pleino (BCP) \cite{Baumgratz2014}. By following it, a number of coherence measures, such as the $l_1$ norm of coherence, the relative entropy of coherence \cite{Baumgratz2014}, and the coherence of formation \cite{Ma2015,Winter2016}, were proposed. Another framework for quantifying coherence is based on translationally invariant operations, put forward by Marvian and Spekkens (MS) in Ref. \cite{Marvian2014}. By following the MS framework, the skew information \cite{Marvian2014,Girolami2014}, the trace norm of commutator \cite{Marvian2014}, and the quantum Fisher information \cite{Yadin2016} were proposed. Besides, based on different free operations \cite{Aberg2006,Yadin.etal2015,Chitambar201602,Chitambar.Gour2016,Streltsov2017},
there are other attempts to quantify coherence, and more coherence measures can be found accordingly.

While various coherence measures have been proposed in theory, an important issue following is how to estimate these coherence measures in experiments. In practical applications, the state of a system is often unknown, and the accessible experimental measurements are typically limited. To estimate a coherence measure of an unknown state from only limited experimental data, the challenge is to establish linkages between the coherence measure and the experimental data available. Furthermore, a coherence measure describes one ability of a state to perform quantum information processing tasks. It is the lower bound of a coherence measure that determines whether a state is qualified for some task. Therefore, the further challenge is to derive the greatest lower bound of a coherent measure of an unknown state rather than a general bound or other approximations.

In this Letter, we put forward an approach to estimate coherence measures in experiments. Our approach, consisting of a basic formula, a computation-friendly formula, and a numerical method, provides a comprehensive way to obtain the greatest lower bounds of coherence measures from measured expectation values of any Hermitian operators.

We use $\rho$ to denote the state of a quantum system, which is unknown. What we know about the state is only the  measurements $ \Tr(\rho O_k)$ of $N$ Hermitian operators $O_1,\dots,O_N$.
Without loss of generality, we assume the experimental data are described as
\begin{eqnarray}\label{Edata}
a_k\leq\Tr(\rho O_k)\leq b_k,~~k=1,\dots,N,
\end{eqnarray}
where $a_k$ and $b_k$ are real numbers \cite{note-Guhne2007}. Let $C(\rho)$ represent a coherence measure, which may be any one of the coherence measures based on the BCP framework, the MS framework, or any other frameworks. $C(\rho)$ is  required  only to be continuous and convex.
We aim to obtain the greatest lower bound of $C(\rho)$ over the states that are compatible with the experimental data, i.e., the infimum,
\begin{eqnarray}\label{BLB}
C^{\textrm{LB}}=\inf\{C(\rho)~|~ a_k\leq\Tr(\rho O_k)\leq b_k,\forall k\}.
\end{eqnarray}

A direct approach of calculating $C^{\textrm{LB}}$ might be to find all the states satisfying Eq. (\ref{Edata}) and identify the least one from all $C(\rho)$. However, such an approach is unfeasible, in general. On one hand, it is difficult to find all the states satisfying Eq. (\ref{Edata}), especially when the experimental data available are very limited. On the other hand, a large family of coherence measures are convex roof measures, and it is already quite difficult to calculate a convex roof coherence measure even for a given mixed state and must be more difficult to estimate the measure for an unknown state.  Here, we will develop an innovative approach of calculating $C^{\textrm{LB}}$, which can steer clear of these difficulties.

{\it First, we establish the basic formula of obtaining $C^{\textrm{LB}}$}. Our idea is to transform the problem of finding the infimum of $C(\rho)$ over the states satisfying the constraints in Eq. (\ref{Edata}) into the optimization problem of a Lagrangian function over the whole state space.

We define the Lagrangian function as
\begin{eqnarray}\label{Lagrangian function}
\mathcal{L}(\rho,\mu,\nu)=C(\rho)+\sum_{k=1}^N\left\{\mu_k\left[-\Tr(\rho O_k)+a_k\right]+\nu_k\left[\Tr(\rho O_k)-b_k\right]\right\},\nonumber\\
\end{eqnarray}
where $\mu:=(\mu_1,\dots,\mu_N)$ and $\nu:=(\nu_1,\dots,\nu_N)$  are two $N$-tuples of real numbers, known as the Lagrange multipliers.  Here, the penalty terms $[-\Tr(\rho O_k)+a_k]$ and $[\Tr(\rho O_k)-b_k]$ originate from the constraints in Eq. (\ref{Edata}).

To find a formula of calculating $C^{\textrm{LB}}$, we first examine the supremum of $\mathcal{L}(\rho,\mu,\nu)$ over $(\mu,\nu)\geq 0$, denoted as $\sup_{(\mu,\nu)\geq 0}\mathcal{L}(\rho,\mu,\nu)$. Here, $(\mu,\nu)\geq 0$ is $\mu_k\geq 0$ and $\nu_k\geq 0$ for all $k$. For convenience, we use $S_c$ to denote the set of states that satisfy the constraints in Eq. (\ref{Edata}). If $\rho\in S_c$, all the penalty terms $[-\Tr(\rho O_k)+a_k]$ and $[\Tr(\rho O_k)-b_k]$ must be nonpositive, and, in this case, the supremum is attained at the point $(\mu,\nu)=0$. Hence, $\sup_{(\mu,\nu)\geq 0}\mathcal{L}(\rho,\mu,\nu)=C(\rho)$ for $\rho\in S_c$. If $\rho\notin S_c$, at least one of the penalty terms is positive, and in the latter case, $\mathcal{L}(\rho,\mu,\nu)$ can be arbitrarily large as long as the Lagrange multiplier associated with the positive penalty term is large enough. Hence,  $\sup_{(\mu,\nu)\geq 0}\mathcal{L}(\rho,\mu,\nu)=+\infty$ for $\rho\notin S_c$.

We then calculate the infimum of $\sup_{(\mu,\nu)\geq 0}\mathcal{L}(\rho,\mu,\nu)$ over all states $\rho$, denoted as  $\inf_{\rho}\sup_{(\mu,\nu)\geq 0}\mathcal{L}(\rho,\mu,\nu)$. We have shown that $\sup_{(\mu,\nu)\geq 0}\mathcal{L}(\rho,\mu,\nu)=C(\rho)$ if $\rho\in S_c$ and $+\infty$ if $\rho\notin S_c$, from which it follows that $\inf_{\rho}\sup_{(\mu,\nu)\geq 0}\mathcal{L}(\rho,\mu,\nu)=\inf_{\rho\in S_c}\sup_{(\mu,\nu)\geq 0}\mathcal{L}(\rho,\mu,\nu)=\inf_{\rho\in S_c}C(\rho)=C^{\textrm{LB}}$. That is,
\begin{eqnarray}\label{inf_sup}
C^{\textrm{LB}}=\inf_{\rho}\sup_{(\mu,\nu)\geq 0}\mathcal{L}(\rho,\mu,\nu).
\end{eqnarray}

Furthermore, $\mathcal{L}(\rho,\mu,\nu)$ is a continuous convex function of $\rho$  and a continuous concave function of $(\mu,\nu)$, where all $\rho$ comprise a compact convex set and all $(\mu,\nu)\geq 0$ comprise a convex set. That is, $\mathcal{L}(\rho,\mu,\nu)$ fulfills the conditions of Sion's minimax theorem \cite{Sion}. We can therefore interchange ``$\inf_{\rho}$'' and ``$\sup_{(\mu,\nu)\geq 0}$'' in Eq. (\ref{inf_sup}) without affecting the result. We then arrive at the formula
\begin{eqnarray}\label{formula 1}
C^{\textrm{LB}}=\sup_{(\mu,\nu)\geq 0}\inf_{\rho}\mathcal{L}(\rho,\mu,\nu).
\end{eqnarray}

Thus, we have transformed the problem of finding the greatest lower bound of $C(\rho)$ subject to the inequality constraints in Eq. (\ref{Edata}) into the optimization problem of a Lagrangian function over the whole state space. The key point of our idea is to convert each constraint $a_k\leq\Tr(\rho O_k)\leq b_k$ into two penalty terms $[-\Tr(\rho O_k)+a_k]$ and $[\Tr(\rho O_k)-b_k]$. With these penalty terms added to $C(\rho)$, the constraints appearing in expression (\ref{BLB}) are removed from formula (\ref{formula 1}). Such a technique can be regarded as an extension of the widely used Lagrange-multiplier technique of maximizing or minimizing a function subject to equality constraints \cite{Boyd}.

Formula (\ref{formula 1}) can be simplified when applied to convex roof coherence measures. It is easy to verify that, in the case of $C(\rho)$ being a convex roof coherence measure, the formula reduces to
\begin{eqnarray}\label{simplified version}
C^{\textrm{LB}}=\sup_{(\mu,\nu)\geq 0}\inf_{\ket{\psi}}\mathcal{L}(\ket{\psi},\mu,\nu),
\end{eqnarray}
where the infimum is taken only over all pure states. In fact, a convex roof coherence measure is defined as  $C(\rho)=\inf_{\{p_k,\ket{\psi_k}\}}\sum_kp_kC(\ket{\psi_k})$, where the infimum is taken over all pure state decompositions of $\rho=\sum_kp_k\ket{\psi_k}\bra{\psi_k}$.  Substituting $C(\rho)=\inf_{\{p_k,\ket{\psi_k}\}}\sum_kp_kC(\ket{\psi_k})$ into Eq. (\ref{Lagrangian function}), we have  $\mathcal{L}(\rho,\mu,\nu)=\inf_{\{p_k,\ket{\psi_k}\}}\sum_kp_k\mathcal{L}(\ket{\psi_k},\mu,\nu)$, which leads to  $C^{\textrm{LB}}=\sup_{(\mu,\nu)\geq 0}\inf_\rho\inf_{\{p_k,\ket{\psi_k}\}}\sum_kp_k\mathcal{L}(\ket{\psi_k},\mu,\nu)$.
Noting that $\inf_\rho\inf_{\{p_k,\ket{\psi_k}\}}\cdot=\inf_{\{\ket{\psi_k}\}}\inf_{\{p_k\}}\cdot$,
we obtain $C^{\textrm{LB}}=\sup_{(\mu,\nu)\geq 0}\inf_{\{\ket{\psi_k}\}}\inf_{\{p_k\}}\sum_kp_k\mathcal{L}(\ket{\psi_k},\mu,\nu)$. Furthermore, since any convex combination of $\{\mathcal{L}(\ket{\psi_k},\mu,\nu)\}$ cannot be less than the least one of them, we finally obtain Eq. (\ref{simplified version}). This formula steers clear of the difficulty of convex roof coherence measures lacking a closed-form expression for mixed states.

Equation (\ref{formula 1}), with its simplified form (\ref{simplified version}) for convex roof coherence measures, provides a basic formula for calculating $C^{\textrm{LB}}$. According to the formula, one can obtain the greatest lower bound of $C(\rho)$ only by calculating $\sup_{(\mu,\nu)\geq 0}\inf_{\rho}\mathcal{L}(\rho,\mu,\nu)$. Examples of showing its usefulness can be seen in the end of this Letter.

{\it Second, we derive a computation-friendly formula of calculating $C^{\textrm{LB}}$}.
Although $C^{\textrm{LB}}$ may be obtained by directly using our formula (\ref{formula 1}) in some cases, one may need to resort to a numerical computation to calculate it in most cases due to the complexity of the problem. Since $\inf_{\rho}\mathcal{L}(\rho,\mu,\nu)$ is a nondifferentiable function of $(\mu,\nu)$, formula (\ref{formula 1}) is not convenient for numerical computation. We now convert formula (\ref{formula 1}) into a computation-friendly expression such that $C^{\textrm{LB}}$ can be figured out numerically.

We first introduce the notion of $\delta$-dependent sample set. A $\delta$-dependent sample set, denoted as $S(\delta)$, is defined as any finite set of states satisfying the condition that, for every state $\rho$, there always exists a state $\rho^\prime\in S(\delta)$ such that $D(\rho,\rho^\prime)<\delta$. Here, $D$ can be any reasonable distance, e.g., the trace distance, and $\delta$ represents a positive number. Because of the compactness of the state space, such a sample set always exists no matter what the value of $\delta$ is. With the notion of $\delta$-dependent sample set, we can define a new function as
\begin{eqnarray}\label{FT}
\mathcal{F}_{S(\delta),T}(\mu,\nu)=-\frac{1}{T}\ln\sum_{\rho\in S(\delta)} \exp[-T\mathcal{L}(\rho,\mu,\nu)],
\end{eqnarray}
where $T$ is a positive number.

We now establish the relation between $\mathcal{F}_{S(\delta),T}(\mu,\nu)$ and $C^{\textrm{LB}}$. To this end, we examine the two functions  $\inf_\rho\mathcal{L}(\rho,\mu,\nu)$ and $\min_{\rho\in S(\delta)}\mathcal{L}(\rho,\mu,\nu)$. Since both of them go to the negative infinity as any $\mu_k$ or $\nu_k$ becomes positively infinite, the maximum values of the functions over $(\mu,\nu)\geq 0$ are achieved at finite values of $(\mu,\nu)$.
Hence, there must exist a compact set of  $(\mu,\nu)$, denoted as $\Omega$, which contains the maximum points of the two functions. Since any continuous function from a compact metric space into a metric space is uniformly continuous \cite{Rudin}, $\mathcal{L}(\rho,\mu,\nu)$ is uniformly continuous over the Cartesian product of the whole state space and $\Omega$. Consequently, for every $\epsilon>0$, there exists $\delta>0$ such that  $\abs{\mathcal{L}(\rho,\mu,\nu)-\mathcal{L}(\rho^\prime,\mu,\nu)}<\epsilon$ for all $(\mu,\nu)\in \Omega$ as long as $D(\rho,\rho^\prime)<\delta$.

According to the definition of $S(\delta)$, for every state $\rho$, there always exists a state $\rho^\prime\in S(\delta)$ satisfying $D(\rho,\rho^\prime)<\delta$. We then have $\abs{\inf_\rho\mathcal{L}(\rho,\mu,\nu)-\min_{\rho\in S(\delta)}\mathcal{L}(\rho,\mu,\nu)}<\epsilon$ for all $(\mu,\nu)\in\Omega$, which further leads to
$\abs{\sup_{(\mu,\nu)\in\Omega}\inf_{\rho}\mathcal{L}(\rho,\mu,\nu)-\sup_{(\mu,\nu)\in\Omega}\min_{\rho\in S(\delta)}\mathcal{L}(\rho,\mu,\nu)}<\epsilon$. As $\Omega$ contains the maximum points, we can here replace $(\mu,\nu)\in\Omega$ with $(\mu,\nu)\geq 0$. Hence, we obtain
\begin{eqnarray}\label{step1}
\lim_{\delta\rightarrow 0}\max_{(\mu,\nu)\geq 0}\min_{\rho\in S(\delta)}\mathcal{L}(\rho,\mu,\nu)=\sup_{(\mu,\nu)\geq 0}\inf_{\rho}\mathcal{L}(\rho,\mu,\nu)=C^{\textrm{LB}}.
\end{eqnarray}

On the other hand, Eq. (\ref{FT}) can be rewritten as
$\mathcal{F}_{S(\delta),T}(\mu,\nu)=\min_{\rho\in S(\delta)}\mathcal{L}(\rho,\mu,\nu)+\Delta$ with  $\Delta=-\frac{1}{T}\ln\sum_{\rho\in S(\delta)}\exp\{-T[\mathcal{L}(\rho,\mu,\nu)-\min_{\rho\in S(\delta)}\mathcal{L}(\rho,\mu,\nu)]\}$. Since $\lim_{T\rightarrow +\infty}\Delta=0$, we immediately have $\lim_{T\rightarrow +\infty}\mathcal{F}_{S(\delta),T}(\mu,\nu)= \min_{\rho\in S(\delta)}\mathcal{L}(\rho,\mu,\nu)$, which further leads to
\begin{eqnarray}\label{tong1}
\lim_{\delta\rightarrow 0}\max_{(\mu,\nu)\geq 0}\lim_{T\rightarrow +\infty}\mathcal{F}_{S(\delta),T}(\mu,\nu)= \lim_{\delta\rightarrow 0}\max_{(\mu,\nu)\geq 0}\min_{\rho\in S(\delta)}\mathcal{L}(\rho,\mu,\nu).
\end{eqnarray}

From Eqs. (\ref{step1}) and (\ref{tong1}), it follows that
\begin{eqnarray}\label{step2}
C^{\textrm{LB}}=\lim_{\delta\rightarrow 0}\max_{(\mu,\nu)\geq 0}\lim_{T\rightarrow +\infty}\mathcal{F}_{S(\delta),T}(\mu,\nu).
\end{eqnarray}
The two operations ``$\max_{(\mu,\nu)\geq 0}$'' and ``$\lim_{T\rightarrow +\infty}$'' in Eq. (\ref{step2}) can be interchanged due to the uniform convergence of $\mathcal{F}_{S(\delta),T}(\mu,\nu)$ \cite{Rudin}. We finally arrive at the desired expression:
\begin{eqnarray}\label{formula 2}
C^{\textrm{LB}}=\lim_{\delta\rightarrow 0}\lim_{T\rightarrow +\infty}\max_{(\mu,\nu)\geq 0}\mathcal{F}_{S(\delta),T}(\mu,\nu),
\end{eqnarray}
referred to as the computation-friendly formula. $\mathcal{F}_{S(\delta),T}(\mu,\nu)$  is a differentiable function of $(\mu,\nu)$, different from $\inf_{\rho}\mathcal{L}(\rho,\mu,\nu)$, and therefore it is much easier to numerically calculate $\max_{(\mu,\nu)\geq 0}\mathcal{F}_{S(\delta),T}(\mu,\nu)$ than  $\sup_{(\mu,\nu)\geq 0}\inf_{\rho}\mathcal{L}(\rho,\mu,\nu)$.

{\it Third, we provide a numerical method of calculating $C^{\textrm{LB}}$}.
Formula  (\ref{formula 2}) shows that the maximum value of $\mathcal{F}_{S(\delta),T}(\mu,\nu)$ over $(\mu,\nu)\geq 0$ converges to $C^{\textrm{LB}}$  as $\delta$ approaches zero and $T$ approaches positive infinity. This indicates that we can obtain an approximation of $C^{\textrm{LB}}$ by calculating $\max_{(\mu,\nu)\geq 0}\mathcal{F}_{S(\delta),T}(\mu,\nu)$ with a sufficiently small $\delta$ and a sufficiently large $T$, and the approximation will be arbitrarily close to its exact value as long as $\delta$ is small enough and $T$ is large enough.
Note that $\delta$ by its very nature represents the state density of the sample set. A small $\delta$ corresponds to a large sample set, i.e., a sample set with a large state density. Hence, we may realize a small $\delta$ by randomly choosing a large number of states comprising the sample set.  The value of $\lim_{T\rightarrow +\infty}\max_{(\mu,\nu)\geq 0}\mathcal{F}_{S(\delta),T}(\mu,\nu)$ will be closer and closer to $C^{\textrm{LB}}$ as more and more states are added to the sample set.

Based on this idea, we suggest a numerical method of calculating $C^{\textrm{LB}}$ as the following steps.\\
(i) Choose an initial sample set of $L_1$ states denoted as $S_1$ and an initial positive number denoted as $T_1$, and calculate the maximum value of $\mathcal{F}_{S_1,T_1}(\mu,\nu)$ over $(\mu,\nu)\geq 0$ \cite{Tong add 1}.\\
(ii) Enlarge $S_1$ by adding $\Delta L_1$ new states such that the new sample set $S_2$ contains $L_2~(=L_1+\Delta L_1)$ states, increase $T_1$ by adding a positive number $\Delta T_1$ such that $T_2=T_1+\Delta T_1$, and calculate the maximum value of $\mathcal{F}_{S_2,T_2}(\mu,\nu)$ over $(\mu,\nu)\geq 0$.\\
(iii) Repeat the above procedure by adding $\Delta L_k$ new states to $S_k$ and a positive number $\Delta T_k$ to $T_k$ until we find $k_0$ such that $\abs{\max_{(\mu,\nu)\geq 0}\mathcal{F}_{S_{k\geq k_0+1},T_{k\geq k_0+1}}(\mu,\nu)-\max_{(\mu,\nu)\geq 0}\mathcal{F}_{S_{k_0},T_{k_0}}(\mu,\nu)}<\varepsilon$  for arbitrarily large $L_{k\geq k_0+1}>L_{k_0}$ and $T_{k\geq k_0+1}>T_{k_0}$, where $\varepsilon$ is a desired tolerance. Then, we can take  $C^{\textrm{LB}}=\max_{(\mu,\nu)\geq 0}\mathcal{F}_{S_{k_0},T_{k_0}}(\mu,\nu)$. Here, the tolerance $\varepsilon$ can be arbitrarily small, depending on the accuracy needed.

It is worth noting that all the above discussions, including Eqs. (\ref{FT}) and  (\ref{formula 2}), are also applicable to convex roof coherence measures. The only difference is that all the sample sets mentioned above comprise of only pure states when they are used for estimating convex roof coherence measures.

{\it Finally, we present examples to show the usefulness of our approach}.
So far, we have put forward a basic formula and a numerical method to estimate coherence measures from measurement values of any Hermitian operators. Here, the operators may be those observables that have been used as coherence or entanglement witnesses \cite{Napoli2016,Horodecki1996} in previous works  or any other available observables. In contrast with the witness technique, which can identify a coherent state from incoherent states but cannot tell the amount of coherence,  our approach can quantitatively estimate the amount of coherence.  We now present examples to show the usefulness of the basic formula and the numerical method.

The physical model is based on a real experiment, performed with a pair of polarized photons \cite{Barbieri2003}. For convenience, we use $\ket{1}$, $\ket{2}$, $\ket{3}$, and $\ket{4}$ to represent the basis $\ket{HH}$, $\ket{HV}$, $\ket{VH}$, and $\ket{VV}$, respectively, where $H$ and $V$ denote the horizontal and vertical polarizations, respectively. In that experiment, the Hermitian operator measured is $O=\ket{\Psi}\bra{\Psi}$ with $\ket{\Psi}=\frac{1}{2}\sum_{i=1}^{4}\ket{i}$, and the expectation value was given as $\Tr(\rho O)=0.0101\pm 0.0013$, i.e., $0.0088\leq\Tr(\rho O)\leq 0.0114$. We aim to calculate the greatest lower bounds of coherence measures based on the only experimental data. Without loss of generality, we consider two coherence measures: the $l_1$ norm of coherence $C_{l_1}$ and the geometric measure of coherence $C_g$. By definition, $C_{l_1}(\rho)=\sum_{i\neq j}\abs{\rho_{ij}}$ with $\rho_{ij}$ being the elements of $\rho$ \cite{Baumgratz2014}, and $C_g(\rho)=\inf_{\{p_k,\ket{\psi_k}\}}\sum_{k}p_kC_g(\ket{\psi_k})$ with $C_g(\ket{\psi})=1-\max_{i}\abs{\langle i|\psi \rangle}^2$ \cite{Streltsov2015}.

We first use our formula (\ref{formula 1}) and its simplified form (\ref{simplified version}) to analytically resolve $C_{l_1}^{\textrm{LB}}$ and $C_g^{\textrm{LB}}$.  To work out $C_{l_1}^{\textrm{LB}}$ analytically, we use formula (\ref{formula 1}). For this model, the Lagrangian function defined in Eq. (\ref{Lagrangian function}) reads $\mathcal{L}_{l_1}(\rho,\mu,\nu)= C_{l_1}(\rho)+\mu[-\Tr(\rho O)+0.0088]+\nu[\Tr(\rho O)-0.0114]$.  Substituting it into formula (\ref{formula 1}), we can obtain $C_{l_1}^{\textrm{LB}}=\sup_{(\mu,\nu)\geq 0}\inf_{\rho}\mathcal{L}_{l_1}(\rho,\mu,\nu)=0.9544$.  To work out $C_g^{\textrm{LB}}$, we need to use formula (\ref{simplified version}), which is specially applicable to convex roof coherence measures. For this model, there is $\mathcal{L}_g(\ket{\psi},\mu,\nu)= C_g(\ket{\psi})+\mu(-\bra{\psi}O\ket{\psi}+0.0088)+\nu(\bra{\psi}O\ket{\psi}-0.0114)$. We can then obtain $C_{g}^{\textrm{LB}}=\sup_{(\mu,\nu)\geq 0}\inf_{\ket{\psi}}\mathcal{L}_g(\ket{\psi},\mu,\nu)=0.1638$. See the Supplemental Material \cite{Supplemental} for details.

We now use our numerical method to calculate  $C_{l_1}^{\textrm{LB}}$ and $C_g^{\textrm{LB}}$. Without loss of generality, we set the desired tolerance to $\varepsilon=0.0001$.

\begin{figure}[htbp]
\centering
\subfigure{
    \includegraphics[width=0.35\textwidth]{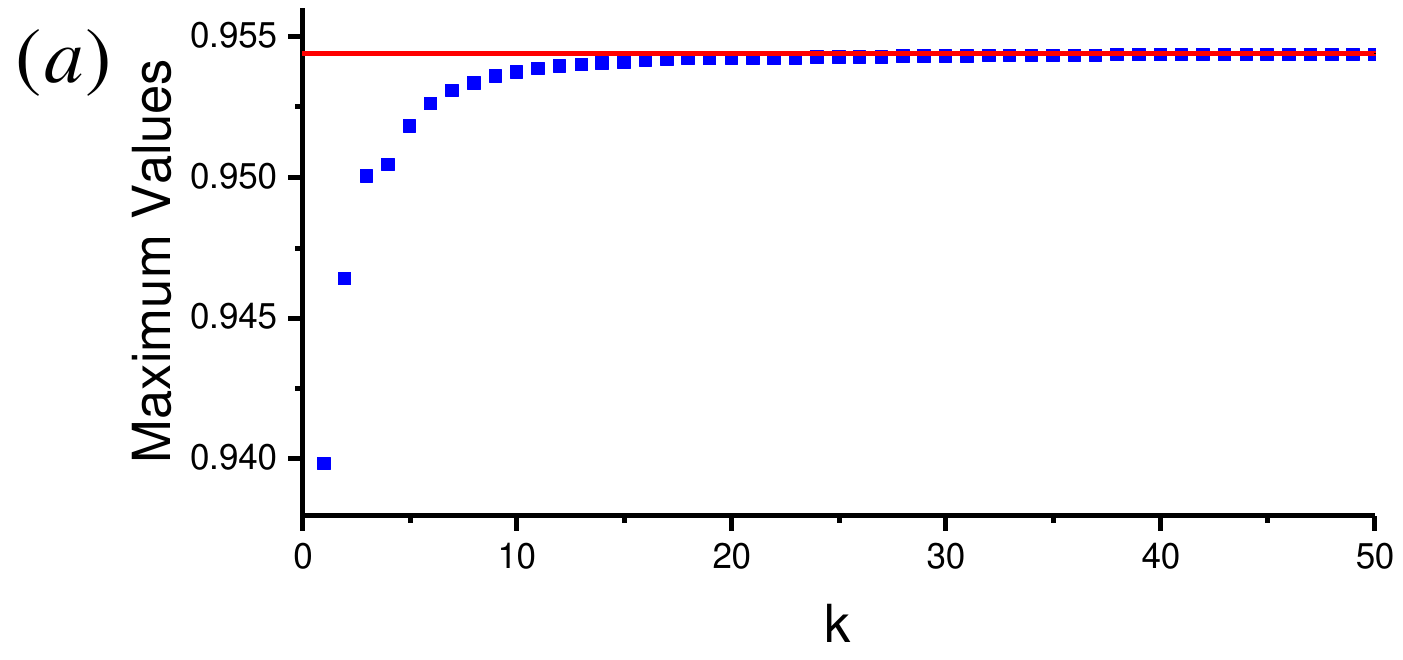}\label{fig1}}
\subfigure{
    \includegraphics[width=0.35\textwidth]{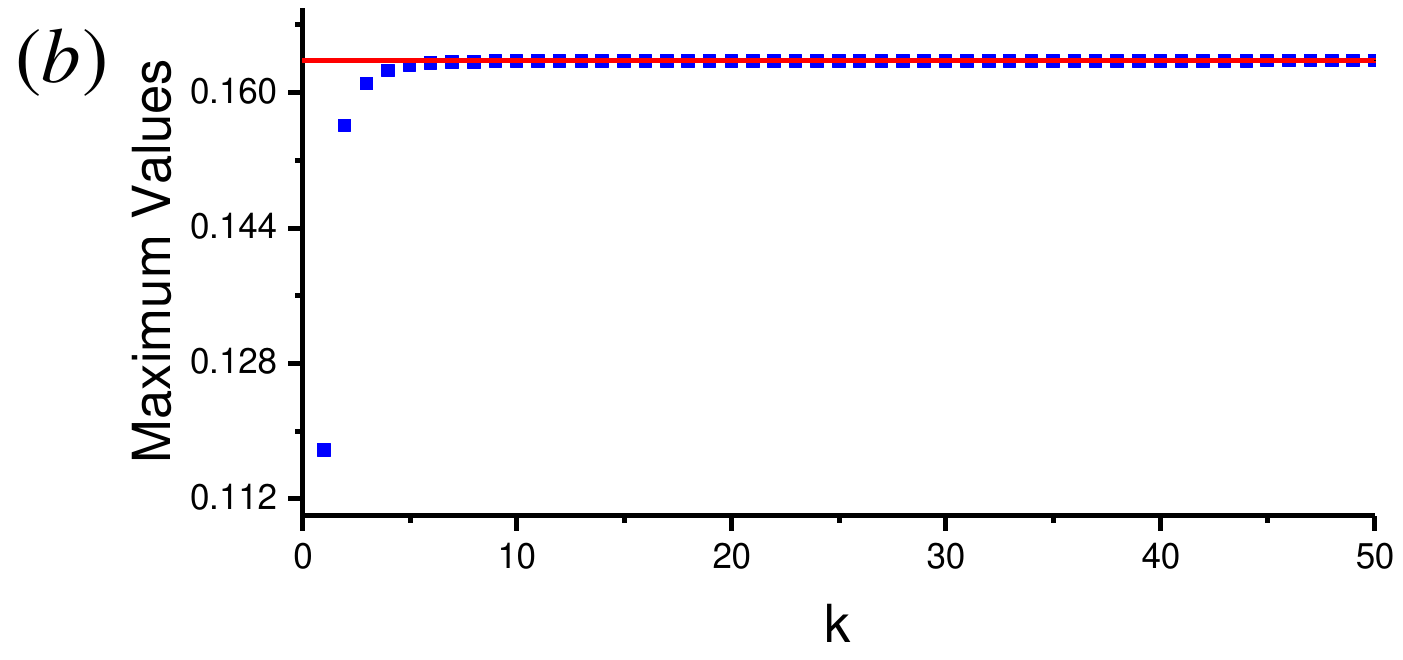}\label{fig2}}
\caption{Plots of $\max_{(\mu,\nu)\geq 0}\mathcal{F}_{S_k,T_k}(\mu,\nu)$ with $S_k$ containing $L_k=100k$ states and $T_k=20k^2$: (a) for $C_{l_1}$ and (b) for $C_g$. Solid lines represent the analytical results. The plots illustrate that $\max_{(\mu,\nu)\geq 0}\mathcal{F}_{S_k,T_k}(\mu,\nu)$ becomes closer and closer to $C^{\textrm{LB}}$ as $L_k$ and $T_k$ increase.}
\end{figure}

To calculate $C_{l_1}^{\textrm{LB}}$, we choose the initial sample set $S_1$ consisting of $L_1=100$ states, which are obtained by randomly generating $100$ density matrices, and let $T_1=20$. We use the gradient ascent method \cite{Boyd} to calculate the maximum value of $\mathcal{F}_{S_1,T_1}(\mu,\nu)$ over $(\mu,\nu)\geq 0$. We then add $100$ new states to $S_1$ to obtain $S_2$ consisting of $L_2=200$ states, add $60$ to $T_1$ to obtain $T_2=80$, and calculate the maximum value of $\mathcal{F}_{S_2,T_2}(\mu,\nu)$ over $(\mu,\nu)\geq 0$. We further repeat the procedure by adding $100$ new states to $S_k$ and $20(2k+1)$ to $T_k$, until we find $\abs{\max_{(\mu,\nu)\geq 0}\mathcal{F}_{S_{k\geq 27},T_{k\geq 27}}(\mu,\nu)-\max_{(\mu,\nu)\geq 0}\mathcal{F}_{S_{26},T_{26}}(\mu,\nu)}<0.0001$ for $L_{k\geq 27}= 100k$ and $T_{k\geq 27}=20k^2$ with all $k=27,28,\dots,50$. Therefore, we take $C_{l_1}^{\textrm{LB}}=\max_{(\mu,\nu)\geq 0}\mathcal{F}_{S_{26},T_{26}}(\mu,\nu)$, which is $0.9543$. The numerical result is shown in Fig. \ref{fig1}, where the solid line represents the analytical result obtained by directly using formula (\ref{formula 1}).

To calculate $C_g^{\textrm{LB}}$, we again choose $L_k=100k$ and $T_k=20k^2$, being the same as that for $C_{l_1}^{\textrm{LB}}$. The only difference is the sample sets of states. Here, the sample sets consist of only pure states, which can be obtained by randomly generating column matrices. By taking the similar procedure to $C_{l_1}^{\textrm{LB}}$, we obtain $C_{g}^{\textrm{LB}}=\max_{(\mu,\nu)\geq 0}\mathcal{F}_{S_{44},T_{44}}(\mu,\nu)=0.1637$. The numerical result is shown in Fig. \ref{fig2}, where the solid line represents the analytical result obtained by directly using expression (\ref{simplified version}).

Before concluding, we stress that our approach is proposed based on the fact that the experimental data available are very limited such that there is no way to know the actual value of a coherence measure. In this case, it becomes vitally important to know the greatest lower bound of the coherence measure. The deviation of the greatest lower bound from the actual value depends on the experimental data available. Generally speaking, the more knowledge we have about the system, the closer to the actual value the greatest lower bound is.

In conclusion, we have proposed an approach to obtain the greatest lower bounds of coherence measures from available experimental data of any Hermitian operators $O_k$, $a_k\leq\Tr(\rho O_k)\leq b_k$. The main idea of our approach is to transform the problem of finding the least value of a coherence measure over the states constrained by $a_k\leq\Tr(\rho O_k)\leq b_k$ into the optimization problem of a Lagrangian function over the whole state space.

Our main findings include Eqs. (\ref{formula 1}) and (\ref{formula 2}). Equation (\ref{formula 1}), with its simplified form (\ref{simplified version}) for convex roof coherence measures, provides a basic formula for obtaining $C^{\textrm{LB}}$. Equation (\ref{formula 2}) provides a line of numerically calculating $C^{\textrm{LB}}$, and, based on it, we suggest a numerical method of approaching  $C^{\textrm{LB}}$ step by step. To show the usefulness of the basic formula and the numerical method, we have presented two examples to each of them.

In passing, we point out that our approach is based only on the requirement that $C(\rho)$ is a continuous and convex function of $\rho$, and therefore it can be naturally extended to other resource measures.

\begin{acknowledgments}
D.-J. Z. acknowledges support from the National Natural Science Foundation of China through Grant No. 11705105. C. L. L. acknowledges support from the National Natural Science Foundation of China through Grant No. 11775129. X.-D. Y. acknowledges support from the National Natural Science Foundation of China through Grant No. 11575101. D. M. T. acknowledges support from the National Basic Research Program of China through Grant No. 2015CB921004.
\end{acknowledgments}


\onecolumngrid
\clearpage

\renewcommand{\theequation}{\thesubsection S.\arabic{equation}}
\setcounter{equation}{0}

\section*{\large{Estimating Coherence Measures from Limited Experimental Data Available\\
 Supplemental Material}}

To work out $C_{l_1}^{\textrm{LB}}$ analytically by using formula (5), we recall the Lagrangian function defined by Eq. (3) in the main text. For the model under consideration, it reads
\begin{eqnarray}
\mathcal{L}_{l_1}(\rho,\mu,\nu)=C_{l_1}(\rho)+\mu[-\Tr(\rho O)+0.0088]
+\nu[\Tr(\rho O)-0.0114].
\end{eqnarray}
We let $\pi:=(\pi(1),\pi(2),\pi(3),\pi(4))$ be a permutation of $\{1,2,3,4\}$,
and $P_{\pi}:=\sum_{i=1}^4\ket{\pi(i)}\bra{i}$ be an operator associated with
$\pi$. It is easy to verify that $C_{l_1}(P_\pi\rho
P_\pi^\dagger)=C_{l_1}(\rho)$ and $P_\pi^\dagger OP_\pi=O$, which result in
$\mathcal{L}_{l_1}(P_\pi\rho
P_\pi^\dagger,\mu,\nu)=\mathcal{L}_{l_1}(\rho,\mu,\nu)$. This is valid for all
permutations $\pi$ in the symmetric group $S_4$. Since
$\mathcal{L}_{l_1}(\rho,\mu,\nu)$ is a convex function of $\rho$, we then have
\begin{eqnarray}
\mathcal{L}_{l_1}(\rho,\mu,\nu)=\frac{1}{4!}\sum_{\pi\in S_4}\mathcal{L}_{l_1}(P_\pi\rho P_\pi^\dagger,\mu,\nu)\geq \mathcal{L}_{l_1}(\frac{1}{4!}\sum_{\pi\in S_4}P_\pi\rho P_\pi^\dagger,\mu,\nu).
\end{eqnarray}
It follows that
\begin{eqnarray}
\inf_{\rho}\mathcal{L}_{l_1}(\rho,\mu,\nu)= \inf_{\rho}\mathcal{L}_{l_1}(\frac{1}{4!}\sum_{\pi\in S_4}P_\pi\rho P_\pi^\dagger,\mu,\nu).
\end{eqnarray}
Further, since there always exists a permutation $\pi^\prime$ satisfying
$i^\prime=\pi^\prime(i)$ and $j^\prime=\pi^\prime(j)$ for any two pairs $(i,j)$
and $(i^\prime,j^\prime)$ with both $i=j$ and $i^\prime=j^\prime$ or both
$i\neq j$ and $i^\prime\neq j^\prime$, we have, by using the rearrangement
theorem,
\begin{eqnarray}
\bra{i^\prime}(\frac{1}{4!}\sum_{\pi\in S_4}P_\pi\rho
P_\pi^\dagger)\ket{j^\prime}=\bra{i}P_{\pi^\prime}^\dagger(\frac{1}{4!}\sum_{\pi\in
S_4}P_\pi\rho P_\pi^\dagger)P_{\pi^\prime}\ket{j}
=\bra{i}(\frac{1}{4!}\sum_{\pi\in S_4}P_\pi\rho P_\pi^\dagger)\ket{j}.
\end{eqnarray}
That is, all the diagonal (off-diagonal) elements of the symmetrizing states are equal to each other, by which the state set $\{\frac{1}{4!}\sum_{\pi\in S_4}P_\pi\rho P_\pi^\dagger\}$ can be equivalently expressed as $\{(1-p)\frac{I}{4}+p\ket{\Psi}\bra{\Psi}~~|-\frac{1}{3}\leq p\leq 1 \}$. Here, $I$ is the $4\times4$ identity matrix. We then arrive at the expression,
\begin{eqnarray}
\inf_{\rho}\mathcal{L}_{l_1}(\rho,\mu,\nu)= \inf_{-\frac{1}{3}\leq p\leq 1}\mathcal{L}_{l_1}((1-p)\frac{I}{4}+p\ket{\Psi}\bra{\Psi},\mu,\nu).
\end{eqnarray}
Using formula (5) in the main text, we finally obtain
\begin{eqnarray}
  \begin{aligned}
    C_{l_1}^{\textrm{LB}}&=\sup_{(\mu,\nu)\geq
    0}\inf_{\rho}\mathcal{L}_{l_1}(\rho,\mu,\nu)\\
    &=\sup_{(\mu,\nu)\geq 0} \inf_{-\frac{1}{3}\leq p\leq
    1}\mathcal{L}_{l_1}((1-p)\frac{I}{4}+p\ket{\Psi}\bra{\Psi},\mu,\nu)\\
    &=\sup_{(\mu,\nu)\geq 0}\inf_{-\frac{1}{3}\leq p\leq
    1}[3\abs{p}-(0.2412+\frac{3}{4}p)\mu+(0.2386+\frac{3}{4}p)\nu]\\
    &=0.9544.
  \end{aligned}
\end{eqnarray}

To work out $C_g^{\textrm{LB}}$ analytically, we use formula (6) in the main text, which is specially applicable to convex roof coherence measures. For the model under consideration, the Lagrangian function reads
\begin{eqnarray}
\mathcal{L}_g(\ket{\psi},\mu,\nu)=
C_g(\ket{\psi})+\mu(-\bra{\psi}O\ket{\psi}+0.0088)
+\nu(\bra{\psi}O\ket{\psi}-0.0114).
\end{eqnarray}
Substituting  $C_g(\ket{\psi})=1-\max_{i}\abs{\langle i|\psi \rangle}^2$
into $\mathcal{L}_g(\ket{\psi},\mu,\nu)$, we have
\begin{eqnarray}
  \begin{aligned}
    \inf_{\ket{\psi}}\mathcal{L}_g(\ket{\psi},\mu,\nu)=&\inf_{\ket{\psi}}\min_i\{\bra{\psi}\left[(\nu-\mu)O-\ket{i}\bra{i}\right]\ket{\psi}+0.0088\mu
    -0.0114\nu +1\}\\
    =&\min_i\{\inf_{\ket{\psi}}\bra{\psi}\left[(\nu-\mu)O-\ket{i}\bra{i}\right]\ket{\psi}+0.0088\mu-0.0114\nu+1\}.
  \end{aligned}
\end{eqnarray}
Note that
$\inf_{\ket{\psi}}\bra{\psi}\left[(\nu-\mu)O-\ket{i}\bra{i}\right]\ket{\psi}$
is just the smallest eigenvalue of $[(\nu-\mu)O-\ket{i}\bra{i}]$, which is
equal to
$-\frac{1}{2}[\sqrt{(\nu-\mu)^2+\nu-\mu+1}-(\nu-\mu)+1]$. We then arrive at the
expression,
\begin{eqnarray}
\inf_{\ket{\psi}}\mathcal{L}_g(\ket{\psi},\mu,\nu)=
-\frac{1}{2}\sqrt{(\nu-\mu)^2+\nu-\mu+1}-0.4912\mu+0.4886\nu+\frac{1}{2},
\end{eqnarray}
which further leads to
\begin{eqnarray}
  \begin{aligned}
    C_g^{\textrm{LB}}&=\sup_{(\mu,\nu)\geq
    0}\inf_{\ket{\psi}}\mathcal{L}_g(\ket{\psi},\mu,\nu)\\
    &=\sup_{(\mu,\nu)\geq
    0}\{-\frac{1}{2}\sqrt{(\nu-\mu)^2+\nu-\mu+1}-0.4912\mu+0.4886\nu+\frac{1}{2}\}\\
    &=0.1638.
  \end{aligned}
\end{eqnarray}


\begin{thebibliography}{99}
\bibitem{Nielsen2010} M. A. Nielsen and I. L. Chuang, \textit{Quantum Computation and Quantum Information} (Cambridge University Press,  Cambridge, England, 2010).
\bibitem{Shor1997} P. W. Shor, SIAM J. Comput. \textbf{26}, 1484 (1997).
\bibitem{Grover1997} L. K. Grover, Phys. Rev. Lett. \textbf{79}, 325 (1997); G. L. Long, Phys. Rev. A \textbf{64}, 022307 (2001).
\bibitem{Grosshans2001} F. Grosshans and P. Grangier, Phys. Rev. Lett. \textbf{88}, 057902 (2002).
\bibitem{Grosshans2003} F. Grosshans, G. Van Assche, J. Wenger, R. Brouri, N. J. Cerf, and P. Grangier, Nature(London) \textbf{421}, 238 (2003).
\bibitem{Giovannetti2004} V. Giovannetti, S. Lloyd, and L. Maccone, Science \textbf{306}, 1330 (2004).
\bibitem{Demkowicz2014} R. Demkowicz-Dobrza\'{n}ski and L. Maccone, Phys. Rev. Lett. \textbf{113}, 250801 (2014).
\bibitem{Giovannetti2011} V. Giovannetti, S. Lloyd, and L. Maccone, Nature Photonics \textbf{5}, 222 (2011).
\bibitem{Aberg2006} J. {\AA}berg, arXiv:quant-ph/0612146.
\bibitem{Gour2008} G. Gour and R. W. Spekkens, New J. Phys. \textbf{10}, 033023 (2008).
\bibitem{Baumgratz2014} T. Baumgratz, M. Cramer, and M. B. Plenio, Phys. Rev. Lett. \textbf{113}, 140401 (2014).
\bibitem{Levi2014}  F. Levi and F. Mintert, New J. Phys. \textbf{16}, 033007 (2014).
\bibitem{Marvian2014} I. Marvian and R. W. Spekkens, Nat. Commun. \textbf{5}, 3821 (2014).
\bibitem{Aberg2014} J. {\AA}berg, Phys. Rev. Lett. \textbf{113}, 150402 (2014).
\bibitem{Streltsov2015} A. Streltsov, U. Singh, H. S. Dhar, M. N. Bera, and G. Adesso, Phys. Rev. Lett. \textbf{115}, 020403 (2015).
\bibitem{Bromley2015} T. R. Bromley, M. Cianciaruso, and G. Adesso, Phys. Rev. Lett. \textbf{114}, 210401 (2015).
\bibitem{Ma2015} X. Yuan, H. Zhou, Z. Cao, and X. Ma, Phys. Rev. A \textbf{92}, 022124 (2015).
\bibitem{Sun2015} Y. Yao, X. Xiao, L. Ge, and C. P. Sun, Phys. Rev. A \textbf{92}, 022112 (2015).
\bibitem{Du2015} S. Du, Z. Bai, and X. Qi, Quantum Inf. Comput. \textbf{15}, 1307 (2015).
\bibitem{Winter2016} A. Winter and D. Yang, Phys. Rev. Lett. \textbf{116}, 120404 (2016).
\bibitem{Bagan2016} E. Bagan, J. A. Bergou, S. S. Cottrell, and M. Hillery, Phys. Rev. Lett. \textbf{116}, 160406 (2016).
\bibitem{Napoli2016} C. Napoli, T. R. Bromley, M. Cianciaruso, M. Piani, N. Johnston, and G. Adesso, Phys. Rev. Lett. \textbf{116}, 150502 (2016).
\bibitem{Radhakrishnan2016} C. Radhakrishnan, M. Parthasarathy, S. Jambulingam, and T. Byrnes, Phys. Rev. Lett. \textbf{116}, 150504 (2016).
\bibitem{JMa2016} J. Ma, B. Yadin, D. Girolami, V. Vedral, and M. Gu, Phys. Rev. Lett. \textbf{116}, 160407 (2016).
\bibitem{Streltsov201602} A. Streltsov, E. Chitambar, S. Rana, M. N. Bera, A. Winter, and M. Lewenstein, Phys. Rev. Lett. \textbf{116}, 240405 (2016).
\bibitem{Chitambar201601} E. Chitambar, A. Streltsov, S. Rana, M. N. Bera, G. Adesso, and M. Lewenstein, Phys. Rev. Lett. \textbf{116}, 070402 (2016).
\bibitem{Chitambar201603} E. Chitambar and M.-H. Hsieh, Phys. Rev. Lett. \textbf{117}, 020402 (2016).
\bibitem{Chitambar201602} E. Chitambar and G. Gour, Phys. Rev. Lett. \textbf{117}, 030401 (2016).
\bibitem{Yu201601}  X.-D. Yu, D.-J. Zhang, C. L. Liu, and D. M. Tong, Phys. Rev. A \textbf{93}, 060303 (R) (2016).
\bibitem{Yu201602} X.-D. Yu, D.-J. Zhang, G. F. Xu, and D. M. Tong, Phys. Rev. A \textbf{94}, 060302 (R) (2016).
\bibitem{Long2016} T. Ma, M.-J. Zhao, S.-M. Fei, and G.-L. Long, Phys. Rev. A \textbf{94}, 042312 (2016).
\bibitem{Fan2016} Y.-R. Zhang, L.-H. Shao, Y. Li, and H. Fan, Phys. Rev. A  \textbf{93}, 012334 (2016).
\bibitem{Hu2016} X. Hu, Phys. Rev. A \textbf{94}, 012326 (2016).
\bibitem{Guo2017} Y.-T. Wang, J.-S. Tang, Z.-Y. Wei, S. Yu, Z.-J. Ke, X.-Y. Xu, C.-F. Li, and G.-C. Guo, Phys. Rev. Lett. \textbf{118}, 020403 (2017).
\bibitem{Zanardi2017} P. Zanardi, G. Styliaris, and L. Campos Venuti, Phys. Rev. A \textbf{95}, 052307 (2017).
\bibitem{Streltsov201603} A. Streltsov, G. Adesso, and M. B. Plenio, Rev. Mod. Phys. \textbf{89}, 041003 (2017).
\bibitem{Fan2017} M.-L. Hu, X. Hu, J.-C. Wang, Y. Peng, Y.-R. Zhang, and H. Fan, arXiv:1703.01852.
\bibitem{Girolami2014} D. Girolami, Phys. Rev. Lett. \textbf{113}, 170401 (2014).
\bibitem{Yadin2016} B. Yadin and V. Vedral, Phys. Rev. A \textbf{93}, 022122 (2016).
\bibitem {Yadin.etal2015} B. Yadin, J. Ma, D.  Girolami, M. Gu, and V. Vedral, Phys. Rev. X \textbf{6}, 041028 (2016).
\bibitem {Streltsov2017} J. I. de Vicente and A. Streltsov, J. Phys. A \textbf{50}, 045301 (2017).
\bibitem {Chitambar.Gour2016} E. Chitambar and G. Gour, Phys. Rev. A \textbf{94}, 052336 (2016).
\bibitem{note-Guhne2007} The experimental data were described by equalities $\Tr(\rho O_k)=a_k$ in Ref. \cite{Guhne2007}, where a method to estimate  entanglement measures was proposed. We here describe the experimental data by inequalities, which
    naturally include the previous equalities as a special case of $a_k=b_k$.
\bibitem{Guhne2007} O. G\"{u}hne, M. Reimpell, and R. F. Werner, Phys. Rev. Lett. \textbf{98}, 110502 (2007).
\bibitem{Sion} M. Sion, Pacific J. Math. \textbf{8}, 171 (1958).
\bibitem{Boyd} S. Boyd and L. Vandenberghe, \textit{Convex Optimization} (Cambridge University Press, Cambridge, England, 2004).
\bibitem{Rudin} W. Rudin, \textit{Principles of Mathematical Analysis} ( McGraw-Hill, New York, 1976).
\bibitem{Tong add 1} The states comprising a sample set do not necessarily fulfill Eq. (\ref{Edata}), and therefore it is easy to construct a sample set. An effective approach is to randomly generate $L$ matrices, denoted as $A_k$, and take $S=\{A_k A_k^\dagger/\Tr(A_kA_k^\dagger),~k=1,2,\cdots,L\}$. Besides, since $\mathcal{F}_{S,T}(\mu,\nu)$ is differentiable, many standard methods, such as the Newton method and the gradient ascent method \cite{Boyd}, can be used to obtain $\max_{(\mu,\nu)\geq 0}\mathcal{F}_{S,T}(\mu,\nu)$.
\bibitem{Horodecki1996} M. Horodecki, P. Horodecki, and R. Horodecki, Phys. Lett. A \textbf{223}, 1 (1996).
\bibitem{Barbieri2003} M. Barbieri, F. De Martini, G. Di Nepi, P. Mataloni, G. M. D'Ariano, and C. Macchiavello, Phys. Rev. Lett. \textbf{91}, 227901 (2003).
\bibitem{Supplemental} See Supplemental Material for details of analytically calculating $C_{l_1}^\textrm{LB}$ and $C_{g}^\textrm{LB}$.
\end{thebibliography}
\end{document}